\documentclass[prx,nofootinbib]{revtex4}
\usepackage{amsmath,srcltx}
\usepackage{amssymb}  
\usepackage{color}  
\usepackage{graphicx}
\usepackage{feynmp}
\usepackage{float}
\newcommand{\be}{\begin{equation}}
\newcommand{\ee}{\end{equation}}
\newcommand{\ba}{\begin{eqnarray}}
\newcommand{\ea}{\end{eqnarray}}

\def\a{\alpha}
\def\b{\beta}

\def\d{\delta}
\def\e{\epsilon}
\def\ve{\varepsilon}

\def\g{\gamma}
\def\h{\eta}
\def\j{\psi}

\def\l{\lambda}
\def\m{\mu}
\def\n{\nu}

\def\p{\pi}

\def\r{\rho}
\def\s{\sigma}
\def\t{\tau}

\def\x{\xi}

\def\D{\Delta}

\def\P{\Pi}

\def\S{\Sigma}

\def\ca{{\cal A}}

\def\co{{\cal O}}

\def\cs{{\cal S}}

\def\cx{{\cal X}}
\def\cy{{\cal Y}}
\def\cz{{\cal Z}}

\newcommand{\ov}{\overline}

\newcommand{\wt}{\widetilde}
\newcommand{\wh}{\widehat}

\newcommand{\aand}{\;\;\;\mbox{and}\;\;\;}

\newcommand{\pa}{\partial}

\newcommand{\pari}{\stackrel{{P}}\longrightarrow}
\def\sl#1{\rlap{\hbox{$\mskip 1 mu /$}}#1}
\def\Sl#1{\rlap{\hbox{$\mskip 3 mu /$}}#1}

\def\I{\leavevmode\hbox{\small1\kern-3.8pt\normalsize1}}

\begin{document}
\title{Quantum parity conservation in planar quantum electrodynamics}
\author{O.M. Del Cima} \email{oswaldo.delcima@ufv.br} 

\author{D.H.T. Franco} \email{daniel.franco@ufv.br}

\author{L.S. Lima} \email{lazaro.lima@ufv.br}

\author{E.S. Miranda} \email{emerson.s.miranda@ufv.br} 

\affiliation{Universidade Federal de Vi\c cosa (UFV),\\
Departamento de F\'\i sica - Campus Universit\'ario,\\
Avenida Peter Henry Rolfs s/n - 36570-900 - Vi\c cosa - MG - Brazil.}
\affiliation{Ibitipoca Institute of Physics (IbitiPhys),\\
36140-000 - Concei\c c\~ao do Ibitipoca - MG - Brazil.}


\begin{abstract}

Quantum parity conservation is verified at all orders in perturbation theory for a massless parity-even $U(1)\times U(1)$ planar quantum electrodynamics (QED$_3$) model. The presence of two massless fermions requires the Lowenstein-Zimmermann (LZ) subtraction scheme, in the framework of the Bogoliubov-Parasiuk-Hepp-Zimmermann-Lowenstein (BPHZL) renormalization method, in order to subtract the infrared divergences induced by the ultraviolet subtractions at 1- and 2-loops, however thanks to the  superrenormalizability of the model the ultraviolet divergences are bounded up to 2-loops. Finally, it is proved that the BPHZL renormalization method preserves parity for the model taken into consideration, contrary to what happens to the ordinary massless parity-even $U(1)$ QED$_3$.     
  
\end{abstract}
\maketitle

\section{Introduction}
The quantum electrodynamics in three space-time dimensions (QED$_3$) \cite{deser-jackiw-templeton-schonfeld} has been considered as a potential theoretical framework for some condensed matter phenomena, namely high-temperature  superconductivity \cite{high-Tc}, quantum Hall effect \cite{quantum-hall-effect}, graphene \cite{graphene}, topological insulators \cite{topological-insulators} and topological superconductors \cite{topological-superconductors}. Some interesting properties may arise in massless, mixed or massive QED$_3$, as parity violation, anyons, topological gauge fields, superrenormalizability and the appearance of infrared divergences. The ordinary massless $U(1)$ QED$_3$ is infrared and ultraviolet perturbatively finite, parity and infrared anomaly free at all orders \cite{massless-all-orders}, however at 1-loop parity is explicitly broken in the course of Lowenstein-Zimmermann (LZ) infrared subtractions in the Bogoliubov-Parasiuk-Hepp-Zimmermann-Lowenstein (BPHZL) renormalization scheme \cite{massless-1-loop}, signalizing that the 1-loop radiatively induced parity-odd Chern-Simons term to the vacuum-polarization tensor is nothing but a counterterm owing to parity-violating LZ infrared subtractions in the BPHZL program\footnote{In perturbation theory, the proof on the absence of a parity anomaly in massless $U(1)$ QED$_3$ has also been performed by  the Epstein-Glaser renormalization method \cite{epstein-glaser}.}. In the meantime, a fundamental question arises if regardless of model the LZ subtraction scheme necessarily violates parity in three space-time dimensions, more specifically, if whether or not parity is broken at any order throughout the infrared subtraction in the Bogoliubov-Parasiuk-Hepp-Zimmermann-Lowenstein (BPHZL) renormalization procedure. Accordingly, the latter issue is dealt in this work by considering a massless parity-even $U(1)\times U(1)$ Maxwell-Chern-Simons QED$_3$ model \cite{masslessU1U1QED3}, with two massless fermions, $\psi_+$ and $\psi_-$, where the gauge mediating bosons, $A_\mu$ (electromagnetic field) and $a_\mu$ (pseudochiral field), associated to the both $U(1)$ symmetries, are massive through a mixed Chern-Simons term. 

The proof presented in this work is organized as follows. In Section \ref{II} the action of the model is introduced and some useful gamma matrices relations are established. Moreover, in Subsections \ref{IIa} and \ref{IIb}, the continuous and discrete classical symmetries, gauge and parity, the propagators and the interactions Feynman rules are presented, the ultraviolet and infrared power countings are fixed for the model, the 1-loop Feynman graphs are identified and the BPHZL subtraction operator defined. In Subsections \ref{IIc} and \ref{IId}, the 1-loop vacuum-polarization tensor and self-energy graphs are presented, and among those ones, the divergents are renormalized. The 2-loop graphs and their BPHZL analyses are left to Section \ref{III}.

\section{BPHZL: 1-loop}\label{II}
In this section the Bogoliubov-Parasiuk-Hepp-Zimmermann momentum space subtraction scheme (BPHZ) \cite{BPHZ}, which does not use any regularization procedure, is applied to 1-loop vacuum-polarization tensor and self-energy divergent graphs. However, due to the presence of massless fermions, $\psi_+$ and $\psi_-$, the momentum subtraction scheme modified by Lowenstein-Zimmermann (BPHZL) \cite{lz} has to be adopted in order to deal with the infrared (IR) divergences that shall arise in the process of ultraviolet (UV) subtractions. 

The action for the massless parity-even $U(1)\times U(1)$ Maxwell-Chern-Simons QED$_3$ model \cite{masslessU1U1QED3}\footnote{A quantum electrodynamics model describing electron-polaron--electron-polaron scattering and four-fold broken degeneracy of the Landau levels in pristine graphene.}, with the parity and gauge invariant Lowenstein-Zimmermann mass term added, is given by:
\ba
\S^{(s-1)} &=& \int{d^3 x} \bigg\{-\frac{1}{4}F^{\m\n}F_{\m\n} -\frac{1}{4}f^{\m\n}f_{\m\n}+ \m \ve^{\m\a\n}A_\m\pa_\a a_\n + 
i {\ov\j_+} {\Sl D}\j_+ + i {\ov\j_-} {\Sl D}\j_- + \nonumber \\
           && \underbrace{-\,m(s-1){\ov\j_+}\j_+ +m(s-1){\ov\j_-}\j_-}_{ \textrm{\small Lowenstein-Zimmermann mass term}}+\, b \pa^\m A_\m+\frac{\a}{2}b^2+\ov{c}\square c+\pi \pa^\m a_\m+\frac{\b}{2}\pi^2+\ov{\x}\square \x \bigg\}~, \label{action} 
\ea
where ${\Sl D}\j_\pm \!\equiv\!(\sl\pa + ie\Sl{A} \pm ig\sl{a})\j_\pm$, $m$ and $\m$ are mass parameters with mass dimension $1$ and the coupling constants $e$ (electric charge) and $g$ (pseudochiral charge) are dimensionful with mass dimension $\frac{1}{2}$. The field strengths, $F_{\m\n}=\pa_\mu A_\nu - \pa_\n A_\m$ and $f_{\m\n}=\pa_\mu a_\nu - \pa_\n a_\m$, are related to the electromagnetic field ($A_\m$) and the pseudochiral gauge field ($a_\m$), respectively. The Dirac spinors $\j_+$ and $\j_-$ are two kinds of fermions where the $\pm$ subscripts are related to their pseudospin sign \cite{masslessU1U1QED3,Binegar}. Also, the fields $c$ and $\x$ are two kind of ghosts\footnote{It is appropriated to stress that neither the ghosts ($c$ and $\x$) nor the antighosts ($\ov{c}$ and $\ov{\x}$) take part of vacuum-polarization tensor, self energy or vertex function Feynman diagrams at any perturbative order, since they are free quantum fields, thus they decouple.} and, $\ov{c}$ and $\ov{\x}$, the two antighosts, whereas $b$ and $\pi$ are the Lautrup-Nakanishi fields \cite{lautrup-nakanishi} playing the role of Lagrange multiplier fields for the gauge conditions. The adopted gamma matrices are $\g^\m=(\s_z,-i\s_x,i\s_y)$. Finally, the Lowenstein-Zimmermann parameter $s$ lies in the interval $0\le s\le1$ and has the same status of an additional subtraction variable (as the external momentum) in the BPHZL renormalization scheme, in such a way that the massless model \cite{masslessU1U1QED3} is recovered by taking $s=1$ at the end of calculations. Furthermore, some conventions and useful relations that shall be used in subsequent calculations follow:
\ba
&&\h^{\m\n}=\textrm{diag}(+--)~,~~\g^\m\g^\n=\h^{\m\n} {\mathbb I}+i\ve^{\m\n\a}\g_\a~,~~{\rm Tr}\{\g^\m \g^\n\}=2\h^{\m\n}~,~~{\rm Tr}\{\g^\m\g^\n\g^\a\}=2i\ve^{\m\n\a}~, \nonumber\\  
&&{\rm Tr}\{\g^{\m_1}\cdots \g^{\m_n}\}=
  \h^{\m_{n-1}\m_n}{\rm Tr}\{\g^{\m_1}\cdots \g^{\m_{n-2}}\}+i\ve^{\m_{n-1}\m_n\a}
  {\rm Tr}\{\g^{\m_1}\cdots \g^{\m_{n-2}}\g_\a\}~.\label{identities}
\ea
It should be pointed out that the trace (${\rm Tr}$) of product of an even number of gamma matrices does not exhibit the Levi-Civita symbol, on the other hand, the trace of product of an odd number (greater than one) of gamma matrices does.

\subsection{Classical symmetries: BRS and parity}\label{IIa}
The action $\S^{(s-1)}$ (\ref{action}) is invariant under the Becchi-Rouet-Stora (BRS) transformations \cite{brs}:
\ba
&s\j_+=i(c + \x)\j_+~,~~s\ov{\j}_+=-i(c + \x)\ov{\j}_+~;& \nonumber \\
&s\j_-=i(c - \x)\j_-~,~~s\ov{\j}_-=-i(c - \x)\ov{\j}_-~;& \nonumber \\
&\displaystyle sA_\m=-\frac{1}{e}\pa_\m c~,~~s c=0~;~~ \displaystyle sa_\m=-\frac{1}{g}\pa_\m \x~,~~s\x=0~;& \nonumber \\
&\displaystyle s\ov{c}=\frac{b}{e}~,~~sb=0~;~~ \displaystyle s\ov{\x}=\frac{\pi}{g}~,~~s\pi=0~;&   \label{BRS}
\ea
as well as under the parity transformations:
\ba
&& \j_+ \pari \j_+^P=-i\g^1\j_-~,~~ \j_- \pari \j_-^P=-i\g^1\j_+~,~~
\ov\j_+ \pari \ov\j_+^P=i\ov\j_-\g^1~,~~ \ov\j_- \pari \ov\j_-^P=i\ov\j_+\g^1~; \nonumber\\
&& A_\mu \pari A_\mu^P=(A_0,-A_1,A_2)~;~~ \phi \pari \phi^P=\phi~,~~\phi=\{b, c, \ov{c}\}~; \nonumber\\
&& a_\mu \pari a_\mu^P=(-a_0,a_1,-a_2)~;~~ \chi \pari \chi^P=-\chi~,~~\chi=\{\pi, \x, \ov{\x}\}~.\label{parity_transformation}
\ea

The tree-level propagators are obtained by taking the free part of the action $\S^{(s-1)}$ (\ref{action}), {\it i.e.}, by switching off the coupling constants $e$ and $g$, thence the propagators in momenta space read:
 \ba
&& \D^{\m\n}_{AA}(k) = -i\bigg\{\frac{1}{k^2-\mu^2}\left(\h^{\m\n}-\frac{k^\m k^\n}{k^2}\right)+\frac{\a}{k^2}\frac{k^\m k^\n}{k^2}\Bigg\}~,~~           
   \D^{\m\n}_{aa}(k) = -i\Bigg\{\frac{1}{k^2-\mu^2}\left(\h^{\m\n}-\frac{k^\m k^\n}{k^2}\right)+\frac{\b}{k^2}\frac{k^\m k^\n}{k^2}\Bigg\}~, \nonumber\\
&& \D_{Aa}^{\m\n}(k) = \frac{\mu}{k^2(k^2-\mu^2)}\e^{\mu\a\n}k_\a~,~~\D_{Ab}^\m(k) = \D_{a \pi}^\m(k) = \frac{k^\m}{k^2}~,~~ \D_{bb}(k) = \D_{\pi\pi}(k) = 0~,~~\D_{\ov{c}c}(k) = \D_{\ov{\x} \x}(k) = -\frac{i}{k^2}~, \nonumber \\        
&& \D_{++}(k) = i\frac{{\sl k}-m(s-1)}{k^2-m^2(s-1)^2}~,~ \D_{--}(k) = i\frac{{\sl k}+m(s-1)}{k^2-m^2(s-1)^2}~.  \label{propk}             
\ea
Notice that from this point forward, all 1- and 2-loops Feynman graphs calculations will be performed in the Landau gauge, $\alpha=\beta=0$. 

The graphical conventions for the propagators are assumed as below:
\begin{fmffile}{feynmanrules}
\begin{equation}
\Delta_{AA}^{\mu \nu} \equiv
\parbox{40pt}{
\begin{fmfgraph*}(13,10)
\fmfleft{i}
\fmfright{o}
\fmf{photon}{i,o}
\end{fmfgraph*}}\quad,\quad
\Delta_{aa}^{\mu \nu} \equiv
\parbox{40pt}{
\begin{fmfgraph*}(13,10)
\fmfleft{i}
\fmfright{o}
\fmf{gluon}{o,i}
\end{fmfgraph*}}
\quad , \quad
\Delta_{Aa}^{\mu \nu}\equiv
\parbox{40pt}{
\begin{fmfgraph*}(13,10)
\fmfleft{i}
\fmfright{o}
\fmf{photon}{i,v}
\fmf{gluon}{o,v}
\end{fmfgraph*}}
\quad , \quad
\Delta_{\pm \pm}\equiv
\parbox{40pt}{
\begin{fmfgraph*}(13,10)
\fmfleft{i}
\fmfright{o}
\fmf{plain}{i,o}
\end{fmfgraph*}} \quad,
\end{equation}
\end{fmffile}
and the Feynman rules for the interaction vertices are given by:
\begin{fmffile}{vertexrules}
	\begin{equation}
	V_{\pm A^{\mu} \pm} \equiv \quad
	\parbox{40pt}{
	\begin{fmfgraph*}(15,15)
		\fmfbottom{i,o}
		\fmftop{u}
		\fmf{plain}{i,v}
		\fmf{plain}{o,v}
		\fmf{photon}{v,u}
		\fmfv{decor.shape=circle,decor.filled=full,decor.size=1.3thick,label=$ie\gamma^{\mu}$,label.angle=130,label.dist=.2cm}{v}
	\end{fmfgraph*}} \quad \quad , \quad \quad
	V_{\pm a^{\mu} \pm} \equiv \quad \quad
	\parbox{40pt}{
	\begin{fmfgraph*}(15,15)
		\fmfbottom{i,o}
		\fmftop{u}
		\fmf{plain}{i,v}
		\fmf{plain}{o,v}
		\fmf{gluon}{v,u}
		\fmfv{decor.shape=circle,decor.filled=full,decor.size=1.3thick,label=$\pm ig\gamma^{\mu}$,label.angle=150,label.dist=.3cm}{v}
	\end{fmfgraph*}} \quad \quad .
	\end{equation}
\end{fmffile}

\subsection{The BPHZL scheme: power counting, subtraction operator, vacuum-polarization and self-energy}\label{IIb}
For the purpose of renormalizing the ultraviolet (UV) and infrared (IR) divergences of all divergent graphs, UV and IR subtraction degrees have to be fixed, to do so the UV and IR dimensions of all the fields shall be determined firstly. For any propagator $\D_{XY}(k,s)$, the UV ($d$) and IR ($r$) dimensions of the fields, $X$ and $Y$, are defined by means of the asymptotical UV and IR behaviour of the propagator, $d_{XY}$ (for $k,s\rightarrow \infty$) and $r_{XY}$ (for $k,(s-1)\rightarrow 0$), respectively, furthermore the following inequalities hold \cite{BPHZ}:
\be
d_X + d_Y \geq 3 + d_{XY} \aand r_X + r_Y \leq 3 + r_{XY}~, \label{uv-ir}
\ee
where, in the Landau gauge, $\alpha=\beta=0$, the UV ($d$) and IR ($r$) dimensions of all the fields are summarized in the Table \ref{table1}. Thus, by taking into account all previous results, the UV ($d(\g)$) and IR ($r(\g)$) superficial degrees of divergence of a 1-particle irreducible Feynman diagram $\g$ stems:
\be
\bordermatrix{ & \cr & d(\g) \cr
&r(\g)  }
 = 3 - \sum\limits_f 
\bordermatrix{& \cr                   
& d_f \cr
& r_f  } N_f  -
\sum\limits_b
\bordermatrix{ & \cr & d_b \cr
&\frac{3}{2}r_b  } N_b +
\bordermatrix{ & \cr & - \cr
& +  }\frac{1}{2} N_e + 
\bordermatrix{ & \cr & - \cr
& +  } \frac{1}{2}N_g -  N_{Aa} ~, \label{power_counting}
\ee
where $N_f$ and $N_b$ are the numbers of external lines of fermions and bosons, respectively, whereas $N_{Aa}$ is the number of internal lines associated to the mixed propagator $\D_{Aa}$. Also, $N_e$ and $N_g$ are the powers of the coupling constants, $e$ and $g$, in the integral corresponding to the graph $\g$. 

\begin{table}[t]
\begin{center}
\begin{tabular}{|c|c|c|c|c|c|c|c|c|c|c|c|c|}
\hline
    & $\j_+$ & $\j_-$      & $A_\m$       & $a_\m$ & $b$ & $\pi$ &$c$  &${\ov c}$ &  $\x$    & $\bar \x$     & $s$ & $s-1$  \\
\hline
$d$ & 1 & 1 & ${1\over 2}$ & ${1\over 2}$ & ${3\over2}$ & ${3\over 2}$ & 0 & 1& 0 & 1& 1 & 1  \\
\hline
$r$ & 1 & 1 &   1          &   1          &   1         &    1         & 0 & 1& 0 & 1& 0 & 1  \\
\hline
\end{tabular}
\end{center}
\caption[]{UV ($d$) and IR ($r$) dimensions.}\label{table1}
\end{table}

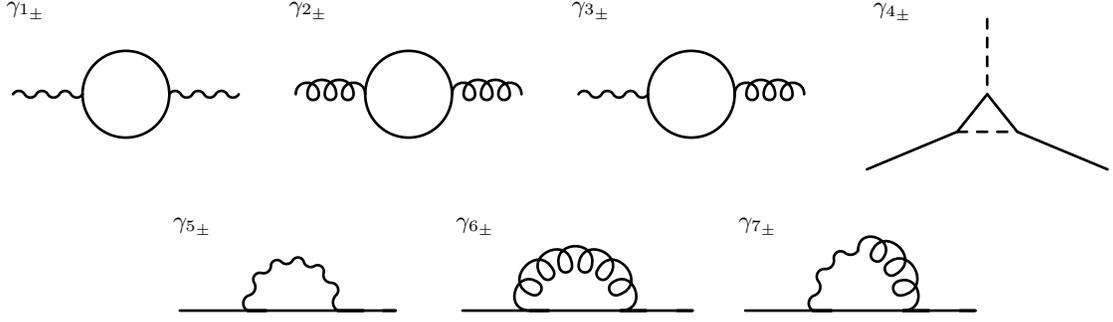
\begin{figure}[H]
\center
\begin{fmffile}{SelfenergyAa}
	\begin{fmfgraph*}(30,10)
		\fmfipair{i,b,c,o}
		\fmfiequ{i}{(0,0)}
		\fmfiv{label=$\gamma_{5_{\pm}}$,label.angle=80,label.dist=1cm}{i}
		\fmfiequ{b}{(.3w,0)}
		\fmfiequ{c}{(.7w,0)}
		\fmfiequ{o}{(.9w,0)}
		\fmfi{photon}{c{up} .. tension .9 .. {down}b}
		\fmfi{plain}{i{right} .. {left}b}
		\fmfi{plain}{b{right} .. {left}c}
		\fmfi{plain}{c{right} .. {left}o}
	\end{fmfgraph*}
	\quad \quad
	\begin{fmfgraph*}(30,10)
		\fmfipair{i,b,c,o}
		\fmfiequ{i}{(0,0)}
		\fmfiv{label=$\gamma_{6_{\pm}}$,label.angle=80,label.dist=1cm}{i}
		\fmfiequ{b}{(.3w,0)}
		\fmfiequ{c}{(.7w,0)}
		\fmfiequ{o}{(.95w,0)}
		\fmfi{gluon}{c{up} .. tension 1 .. {down}b}
		\fmfi{plain}{i{right} .. {left}b}
		\fmfi{plain}{b{right} .. {left}c}
		\fmfi{plain}{c{right} .. {left}o}
	\end{fmfgraph*}
		\quad \quad
	\begin{fmfgraph*}(30,10)
		\fmfipair{i,b,a,c,o}
		\fmfiequ{i}{(0,0)}
		\fmfiv{label=$\gamma_{7_{\pm}}$,label.angle=80,label.dist=1cm}{i}
		\fmfiequ{b}{(.3w,0)}
		\fmfiequ{c}{(.7w,0)}
		\fmfiequ{a}{(.5w,0.8cm)}
		\fmfiequ{o}{(.95w,0)}
		\fmfi{photon}{b{up} .. tension 1 .. {right}a}
		\fmfi{gluon}{c{up} .. tension 1 .. {left}a}
		\fmfi{plain}{i{right} .. {left}b}
		\fmfi{plain}{b{right} .. {left}c}
		\fmfi{plain}{c{right} .. {left}o}
	\end{fmfgraph*}
\end{fmffile}
\caption{The 1-loop diagrams $\gamma_{1_{\pm}}$, $\gamma_{2_{\pm}}$ and $\gamma_{3_{\pm}}$ are the vacuum-polarization tensors, $\gamma_{4_{\pm}}$ is the vertex functions and $\gamma_{5_{\pm}}$, $\gamma_{6_{\pm}}$ and $\gamma_{7_{\pm}}$ are the self-energies. The continuous line represents external legs or propagators of either $\j_+$ or $\j_-$, whereas the dashed lines in $\gamma_{4_{\pm}}$ denote the propagator of $A_\m$, $a_\m$ or the mixed one, and the external leg of either $A_\m$ or $a_\m$.}\label{1loopgraphs}
\end{figure}

The 1-loop vacuum-polarization tensors, self energies and vertex functions diagrams are identified in Fig. \ref{1loopgraphs}, whereas their respectives UV and IR superficial degrees of divergence are displayed in Table \ref{table2}. At this time, it should be mentioned that for any graph $\gamma_{i_{\pm}}$ the subscript $\pm$ refers to external legs or internal lines of either $\j_+$ or $\j_-$.

\begin{figure}[H]
\center
\begin{fmffile}{vacuumpolarization}
	\begin{fmfgraph*}(30,20)
		\fmfleft{i}
		\fmfv{label=$\gamma_{1_{\pm}}$,label.angle=80,label.dist=1cm}{i}
    	\fmfright{o}
    	\fmf{photon}{i,v1}
    	\fmf{photon}{v2,o}
		\fmf{plain,left,tension=0.4}{v1,v2,v1}
	\end{fmfgraph*}
	\quad \quad 
	\begin{fmfgraph*}(30,20)
		\fmfleft{i}
		\fmfv{label=$\gamma_{2_{\pm}}$,label.angle=80,label.dist=1cm}{i}
    	\fmfright{o}
    	\fmf{gluon}{v1,i}
    	\fmf{gluon}{o,v2}
		\fmf{plain,left,tension=0.4}{v1,v2,v1}
	\end{fmfgraph*}
	\quad \quad
	\begin{fmfgraph*}(30,20)
		\fmfleft{i}
		\fmfv{label=$\gamma_{3_{\pm}}$,label.angle=80,label.dist=1cm}{i}
    	\fmfright{o}
    	\fmf{photon}{i,v1}
    	\fmf{gluon}{o,v2}
		\fmf{plain,left,tension=0.4}{v1,v2,v1}
	\end{fmfgraph*}
	\quad
	\begin{fmfgraph*}(40,20)
	    \fmftop{i1}
	    \fmfleft{i2,i4}
	    \fmfv{label=$\gamma_{4_{\pm}}$,label.angle=80,label.dist=2cm}{i2}
	    \fmfright{o,o1}
	    \fmf{dashes}{i1,v1}
	    \fmf{plain}{i2,v2}
	    \fmf{plain}{v3,o}
	    \fmf{plain}{v2,v1}
	    \fmf{plain}{v3,v1}
	    \fmf{dashes}{v2,v3}
	  \end{fmfgraph*}	
\end{fmffile}
\end{figure}

\begin{table}[h]
\begin{center}
\begin{tabular}{|c|c|c|c|c|c|c|c|c|}
\hline
         &$\gamma_{1_{\pm}}$ & $\gamma_{2_{\pm}}$ & $\gamma_{3_{\pm}}$           & $\gamma_{4_{\pm}}^{(a)}$ & $\gamma_{4_{\pm}}^{(b)}$    &  $\gamma_{5_{\pm}}$   & $\gamma_{6_{\pm}}$  &$\gamma_{7_{\pm}}$    \\
\hline
$d$      & 1     &    1   &   1   &  $-1$  & $-2$  & 0   &  0 &  $-1$      \\
\hline
$r$      &  1    &   1    &    1             &   1      &  0    &   2           &    2         &  1        \\
\hline
\end{tabular}
\end{center}
\caption[]{UV ($d$) and the IR ($r$) superficial degrees of divergence of a 1-particle irreducible Feynman diagrams of Fig. \ref{1loopgraphs}, where for the vertex functions $\gamma_{4_{\pm}}^{(a)}$ the dashed internal line represents either $\D_{AA}$ or $\D_{aa}$ propagators, while for $\gamma_{4_{\pm}}^{(b)}$ it symbolizes the mixed propagator $\D_{Aa}$.} \label{table2}
\end{table}

Accordingly to the power counting theorem, in view of the fact that the 1-loop diagrams $\gamma_{1_{\pm}}$, $\gamma_{2_{\pm}}$, $\gamma_{3_{\pm}}$, $\gamma_{5_{\pm}}$
and $\gamma_{6_{\pm}}$ (see Fig. \ref{1loopgraphs} and Table \ref{table2}) are superficially UV divergent, they have to be UV and IR subtracted. Whenever a graph $\g$ is possibly UV divergent, {\it i.e.} $d(\g)\geq 0$, the BPHZL renormalization method is followed so as to make the graph convergent \cite{lz} by also subtracting the IR divergences induced by the UV subtractions. The BPHZL subtraction program consists of performing UV and IR subtraction operations upon a UV divergent Feynman graph integrand, $I_{\gamma}(p,k,s)$:
\begin{equation}\label{R}
	R_{\gamma}(p,k,s)=\left(1-t^{\rho(\gamma)-1}_{p,s-1}\right)
	\left(1-t^{\delta(\gamma)}_{p,s}\right)I_{\gamma}(p,k,s) ~,
\end{equation}
where $R_\g(p,k,s)$ is the renormalized integrand, which is UV convergent. Moreover, $\d(\g)$ and $\r(\g)$ are the UV and IR degrees of subtraction, respectively, given by \cite{lz}: 
\be
\d(\g) = d(\g) + b(\g) \aand \r(\g) = r(\g) -c(\g)~, \label{degrees}
\ee
where at 1-loop $b(\g)$ and $c(\g)$ are non-negative integers constrained as follows:
\be
\r(\g)\leq \d(\g)+1~, \label{ineq}
\ee
with $t^\t_{x,y}$ being the Taylor expansion operator about $x=y=0$ to order $\t$, provided $\t\geq 0$.

\subsection{The vacuum-polarization tensor}\label{IIc}
The BPHZL renormalization procedures of all 1-particle irreducible vacuum-polarization tensor divergent diagrams (see Fig. \ref{1loopgraphs} and Table \ref{table2}) are rather similar, since they possess the same loop structure, their integrands are equal up to coupling constants dependent factors, $\pm e^2$, $\pm g^2$ and $\mp eg$, corresponding to the 1-loop graphs, $\gamma_{1_{\pm}}$, $\gamma_{2_{\pm}}$ and $\gamma_{3_{\pm}}$, respectively. Initially, the analysis is carried out for the $\gamma_{1_{\pm}}$ Feynman graphs, where the 1-loop vacuum-polarization tensor, $\P_{\gamma_{1_{\pm}}}^{\m\n}(p,s)$, reads
\ba
\P_{\gamma_{1_{\pm}}}^{\m\n}(p,s)
=\int{\frac{d^3k}{(2\p)^3}}\underbrace{\left\{-e^2{\rm Tr}\left[\g^\m
\frac{{\sl k}\mp m(s-1)}{k^2-m^2(s-1)^2} \g^\n\frac{{\sl k -
\sl p}\mp m(s-1)}{(k-p)^2-m^2(s-1)^2}\right]\right\}}_{I_{\gamma_{1_{\pm}}}^{\m\n}(p,k,s)}~.
\ea 

Bearing in mind the conditions (\ref{degrees}) and the inequality (\ref{ineq}), by taking $b(\g_{1_{\pm}})=c(\g_{1_{\pm}})=0$, the UV and IR subtraction degrees are such that $\d(\g_{1_{\pm}})=\r(\g_{1_{\pm}})=1$. Consequently, the 1-loop BPHZL subtracted (renormalized) integrand, $R_{\gamma_{1_{\pm}}}^{\m\n}(p,k,s)$, is written in terms of the unsubtracted one, $I_{\gamma_{1_{\pm}}}^{\m\n}(p,k,s)$, in the following way: 
\ba
R_{\gamma_{1_{\pm}}}^{\m\n}(p,k,s)&=&(1-t^{0}_{p,s-1})(1-t^{1}_{p,s})
I_{\gamma_{1_{\pm}}}^{\m\n}(p,k,s)\nonumber\\
&=&(1-t^{0}_{p,s-1}-t^{1}_{p,s}+t^{0}_{p,s-1}t^{1}_{p,s})
I_{\gamma_{1_{\pm}}}^{\m\n}(p,k,s)~.
\ea  
However, as previously mentioned by setting $s=1$ at the end of all Taylor expansion operations, to retrieve the massless condition, the subtracted integrand, $R_{\g_{1_{\pm}}}^{\m\n}(p,k,1)$, results:
\be
R_{\gamma_{1_{\pm}}}^{\m\n}(p,k,1)=\underbrace{I_{\gamma_{1_{\pm}}}^{\m\n}(p,k,1)}_{\rm parity-even}-\underbrace{I_{\gamma_{1_{\pm}}}^{\m\n}(0,k,1)}_{\rm parity-even}-
\underbrace{p^\r\frac{\pa}{\pa p^\r}I_{\gamma_{1_{\pm}}}^{\m\n}(p,k,s)}_{\rm parity-odd}
\Bigg|_{p=s=0}~, \label{Rpk1}
\ee 
where 
\ba
&&I_{\gamma_{1_{\pm}}}^{\m\n}(p,k,1)=-e^2{\rm Tr}~\left\{\g^\m\frac{{\sl k}}{k^2} \g^\n\frac{{\sl k - \sl p}}{(k-p)^2}\right\}~,~~ I_{\gamma_{1_{\pm}}}^{\m\n}(0,k,1)=-e^2{\rm Tr}~\left\{\g^\m\frac{{\sl k}}{k^2} \g^\n\frac{{\sl k}}{k^2}\right\}~, \label{Ipk1}\\
&&p^\r\frac{\pa}{\pa p^\r}I_{\gamma_{1_{\pm}}}^{\m\n}(p,k,s)\Bigg|_{p=s=0}
=-e^2{\rm Tr}~\left\{\g^\m\frac{{\sl k}\mp m}{k^2-m^2} \g^\n
\left[-\frac{\sl p}{k^2-m^2}+2p\cdot k \frac{{\sl k}\mp m}{(k^2-m^2)^2} \right]\right\}~. \label{I0k0}
\ea 
In addition to that, since the renormalized vacuum-polarization tensor, $\P_{\gamma_{1_{\pm}}}^{(R)\m\n}(p,s)$, is defined by  
\begin{equation}
\P_{\gamma_{1_{\pm}}}^{(R)\m\n}(p,s)= \int \frac{d^3k}{(2\p)^3}~R^{\m\n}_{\gamma_{1_{\pm}}}(p,k,s)~,
\end{equation}
and recalling to the fact that the issue here is to verify if Levi-Civita symbol $\epsilon^{\mu\nu\rho}$ dependent terms might be induced by UV and IR subtractions, only parity-odd pieces of the subtracted integrand, $R_{{\rm odd}\gamma_{1_{\pm}}}^{\m\n}(p,k,1)$ (\ref{Rpk1}), shall be taken into account, then from the Eqs.(\ref{Rpk1})--(\ref{I0k0}), leads to 
\begin{equation}\label{vp1}
	\P^{(R)\m\n}_{{\rm odd}\gamma_{1_{\pm}}}=\pm \frac{e^2m}{4\p |m|}\epsilon^{\m\a\n}
	p_{\a}~,
\end{equation}
with $\P^{(R)\m\n}_{{\rm odd}\gamma_{1_{\pm}}}\equiv\P^{(R)\m\n}_{{\rm odd}\gamma_{1_{\pm}}}(p,1)$.

Analogously to the previous case, $\P^{(R)\m\n}_{{\rm odd}\gamma_{1_{\pm}}}$, the renormalized parity-odd vacuum-polarization tensors $\P^{(R)\m\n}_{{\rm odd}\gamma_{2_{\pm}}}$ and $\P^{(R)\m\n}_{{\rm odd}\gamma_{3_{\pm}}}$, corresponding to $\gamma_{2_{\pm}}$ and $\gamma_{3_{\pm}}$ diagrams, are respectively given by 
\begin{equation}\label{vp2}
	\P^{(R)\m\n}_{{\rm odd}\gamma_{2_{\pm}}}=\pm \frac{g^2m}{4\p |m|}\epsilon^{\m\a\n}
	p_{\a} \aand 
	\P^{(R)\m\n}_{{\rm odd}\gamma_{3_{\pm}}}=\mp \frac{egm}{4\p |m|}\epsilon^{\m\a\n}p_{\a}~.
\end{equation}
Finally, the 1-loop renormalized parity-odd vacuum polarization tensors, $\P^{(R)\m\n}_{{\rm odd}\gamma_1}$, $\P^{(R)\m\n}_{{\rm odd}\gamma_2}$ and $\P^{(R)\m\n}_{{\rm odd}\gamma_3}$ :
\begin{equation}
  \P^{(R)\m\n}_{{\rm odd}\gamma_1}=\P^{(R)\m\n}_{{\rm odd}\gamma_{1_{+}}}+\P^{(R)\m\n}_{{\rm odd}\gamma_{1_{-}}}\equiv 0 ~,~~
  \P^{(R)\m\n}_{{\rm odd}\gamma_2}=\P^{(R)\m\n}_{{\rm odd}\gamma_{2_{+}}}+\P^{(R)\m\n}_{{\rm odd}\gamma_{2_{-}}}\equiv 0 ~,~~
  \P^{(R)\m\n}_{{\rm odd}\gamma_3}=\P^{(R)\m\n}_{{\rm odd}\gamma_{3_{+}}}+\P^{(R)\m\n}_{{\rm odd}\gamma_{3_{-}}}\equiv 0~, 
\end{equation}
vanishes identically. In conclusion, besides there is no 1-loop counterterm for the mixed Chern-Simons term, $\ve^{\m\a\n}A_\m\pa_\a a_\n$ -- which sets out that the 1-loop $\b$-function associated to the Chern-Simons mass parameter ($\m$) vanishes -- the BPHZL subtraction scheme applied to the 1-loop vacuum-polarization tensor preserves parity, being the opposite to what takes place in ordinary massless $U(1)$ QED$_3$ \cite{massless-1-loop}.

\subsection{The self-energy}\label{IId}
Among the six self-energy diagrams (see Fig. \ref{1loopgraphs} and Table \ref{table2}), two are UV finite, $\gamma_{7_{\pm}}$, while the four remaining, $\gamma_{5_{\pm}}$ and $\gamma_{6_{\pm}}$, are UV divergent, thus those which have to be renormalized. However, the BPHZL subtraction procedures for the 1-particle irreducible self-energy divergent diagrams are analogous, differing only by coupling constants dependent factors, $\pm e^2$ and $\pm g^2$, corresponding to the 1-loop graphs, $\gamma_{5_{\pm}}$ and $\gamma_{6_{\pm}}$, respectively. Starting the analysis with $\gamma_{5_{\pm}}$ Feynman graphs, the 1-loop self-energy, $\Sigma(\gamma_{5_{\pm}})$, reads 
\begin{equation}
	\Sigma(\gamma_{5_{\pm}})=
	\int\frac{d^3k}{(2\pi)^3}
	\underbrace{\left\{-e^2\gamma^{\mu}\left[\frac{1}{k^2-\mu^2}
	\left(\eta_{\mu \nu}-\frac{k_{\mu}k_{\nu}}{k^2}\right)\right] 
	\left[\frac{(\sl k - \sl p)\mp m(s-1)}{(k-p)^2-m^2(s-1)^2}
	\right]\gamma^{\nu}
	\right\}}_{I_{\gamma_{5_{\pm}}}(p,k,s)}.
\end{equation}

Keeping in mind one more time the conditions (\ref{degrees}) and the inequality (\ref{ineq}), the UV and IR subtraction degrees are $\d(\gamma_{5_{\pm}})=\d(\gamma_{6_{\pm}})=0$ and $\r(\gamma_{6_{\pm}})=\r(\gamma_{5_{\pm}})=1$, where it has been fixed $b(\gamma_{5_{\pm}})=b(\gamma_{6_{\pm}})=0$ and $c(\gamma_{5_{\pm}})=c(\gamma_{6_{\pm}})=1$. The 1-loop BPHZL subtracted (renormalized) integrand, $R_{\gamma_{5_{\pm}}}(p,k,s)$, can be expressed in terms of the unsubtracted one, $I_{\gamma_{5_{\pm}}}(p,k,s)$, as follows:   
\ba
R_{\gamma_{5_{\pm}}}(p,k,s)&=&(1-t^{0}_{p,s-1})(1-t^{0}_{p,s})I_{\gamma_{5_{\pm}}}(p,k,s)\nonumber\\
&=&(1-t^{0}_{p,s-1}-t^{0}_{p,s}+t^{0}_{p,s-1}t^{0}_{p,s})I_{\gamma_{5_{\pm}}}(p,k,s)~.
\ea  
Yet again, setting $s=1$ at the end of the Taylor expansion operations, restoring the massless condition, the subtracted integrand, $R_{\g_{5_{\pm}}}(p,k,1)$, results:
\begin{equation}\label{Rself}
  R_{\g_{5_{\pm}}}(p,k,1)=I_{\g_{5_{\pm}}}(p,k,1)-I_{\gamma_{5_{\pm}}}(0,k,1)~,
\end{equation}
where,  
\be
I_{\g_{5_{\pm}}}(p,k,1) = 2e^2\left\{\frac{1}{k^2-\m^2}\frac{1}{(k-p)^2}  \left[\sl k - \frac{\sl k (k \cdot p)}{k^2} \right]  \right\}~,~~ I_{\gamma_{5_{\pm}}}(0,k,1) = 2e^2 \frac{\sl k}{k^2(k^2-\m^2)}~.\label{Ipk1se}
\ee
Additionally, once the renormalized self-energy, $\S_{\gamma_{5_{\pm}}}^{(R)}(p,s)$, is defined by   
\begin{equation}
\S_{\gamma_{5_{\pm}}}^{(R)}(p,s)=\int \frac{d^3 k}{(2\pi)^3}~R_{\gamma_{5_{\pm}}}(p,k,s),
\end{equation}
such that, from the Eqs.(\ref{Rself})--(\ref{Ipk1se}), leads to
\begin{equation}
	\S^{(R)}_{\gamma_{5_{\pm}}}
	=-\frac{ie^2 \sl p}{4\p}\left[\frac{1}{4\sqrt{p^2}}\left(\frac{p^2}{\m^2}+
	\frac{3\m^2}{p^2}+2\right)
	\ln \left(\frac{\m^2-p^2}{(\sqrt{\m^2}-\sqrt{p^2})^2}\right)
	-\frac{|\m|}{2}\left(\frac{1}{\m^2}+\frac{3}{p^2}\right)
	+i\p\frac{p^2}{4\m^2\sqrt{p^2}}
	 \right]~,
\end{equation}
with $\S^{(R)}_{\gamma_{5_{\pm}}}\equiv\S^{(R)}_{\gamma_{5_{\pm}}}(p,1)$.

Similarly to the previous case, $\S^{(R)}_{\gamma_{5_{\pm}}}$, the renormalized self-energies $\S^{(R)}_{\gamma_{6_{\pm}}}$, corresponding to $\gamma_{6_{\pm}}$ diagram, read 
\begin{equation}
	\S^{(R)}_{\gamma_{6_{\pm}}}
	=-\frac{ig^2 \sl p}{4\p}\left[\frac{1}{4\sqrt{p^2}}\left(\frac{p^2}{\m^2}+
	\frac{3\m^2}{p^2}+2\right)
	\ln \left(\frac{\m^2-p^2}{(\sqrt{\m^2}-\sqrt{p^2})^2}\right)
	-\frac{|\m|}{2}\left(\frac{1}{\m^2}+\frac{3}{p^2}\right)
	+i\p\frac{p^2}{4\m^2\sqrt{p^2}}
	 \right]~.
\end{equation}

Accordingly, the 1-loop renormalized self-energies, $\S^{(R)}_+=\S^{(R)}_{\gamma_{5_+}}+\S^{(R)}_{\gamma_{6_+}}$ and $\S^{(R)}_-=\S^{(R)}_{\gamma_{5_-}}+\S^{(R)}_{\gamma_{6_-}}$, associated respectively to $\j_+$ and $\j_-$:
\ba
	\S^{(R)}_+=\S^{(R)}_-
	&\!\!=\!\!&-\frac{i(e^2+g^2) \sl p}{4\p}\left[\frac{1}{4\sqrt{p^2}}\left(\frac{p^2}{\m^2}+
	\frac{3\m^2}{p^2}+2\right)
	\ln \left(\frac{\m^2-p^2}{(\sqrt{\m^2}-\sqrt{p^2})^2}\right)
	-\frac{|\m|}{2}\left(\frac{1}{\m^2}+\frac{3}{p^2}\right)
	+i\p\frac{p^2}{4\m^2\sqrt{p^2}}
	 \right] \nonumber\\
	 &\!\!=\!\!&\frac{(e^2+g^2)}{4\p} {\sl p}~\co(p^2,\m)~,
\ea
contribute to the 1-loop effective action (in momenta space) with the following term: 
\be
\ov\j_+ \S^{(R)}_+ \j_+ + \ov\j_- \S^{(R)}_-\j_- ~~\overset{P} {\longmapsto}~~ 
\ov\j_- \S^{(R)}_-\j_- + \ov\j_+ \S^{(R)}_+ \j_+~,
\ee
that shows to be invariant under parity, thereby the BPHZL subtraction scheme does not break parity in the case of the 1-loop self-energy either. Finally, it has been finished the proof on the BPHZL parity invariance at 1-loop for the massless parity-even $U(1)\times U(1)$ Maxwell-Chern-Simons QED$_3$ model \cite{masslessU1U1QED3}. Nevertheless, thanks to divergent 2-loops vacuum polarization tensor diagrams (see Fig. \ref{2loopgraphs}), it remains to verify if whether or not parity still be preserved in the course of the 2-loops BPHZL ultraviolet and infrared subtractions.

\section{BPHZL: 2-loops}\label{III}
In order to complete the proof if parity is broken or not by the BPHZL renormalization method, once the model into consideration here is superrenormalizable and the ultraviolet divergences are bounded up to 2-loops (\ref{power_counting}), it still remains to identify and investigate the potential UV divergent 2-loops diagrams in what concerns parity breakdown. By power-counting inspection (\ref{power_counting}), exclusively twenty four of the thirty six vacuum-polarization tensor Feynman graphs\footnote{It should be pointed that, for the sake of subsequent  renormalization, the symmetrical diagrams corresponding to $\gamma_{11_{\pm}}$, $\gamma_{12_{\pm}}$ and $\gamma_{13_{\pm}}$ -- those with the propagators $\D^{\m\n}_{AA}$, $\D^{\m\n}_{aa}$ or $\D^{\m\n}_{Aa}$ inside the loop in its upper part -- have to be taken into consideration.} show to be divergent at 2-loops (see Fig. \ref{2loopgraphs}), furthermore, it shall be verified if parity-odd local counterterms, with UV dimension 2, of the type $\epsilon^{\mu \a \nu}A_{\mu}p_{\a}A_{\nu}$ or $\epsilon^{\mu \a \nu}a_{\mu}p_{\a}a_{\nu}$ -- local counterterm of the type $\epsilon^{\mu \a \nu}A_{\mu}p_{\a}a_{\nu}$ shall be discarded throughout this analysis because it is parity-even -- might be generated by the UV and IR subtractions. However, power-counting (\ref{power_counting}) dimensional analysis reveals that even though parity-odd Levi-Civita symbol dependent counterterms could appear, they would be nonlocal since their coupling constant order should be of mass dimension 2, namely, $e^4$, $e^2g^2$ or $g^4$. 

\begin{figure}[h]
\center
	\begin{fmffile}{twoloopgraphs}
	\begin{fmfgraph*}(37,25)
		\fmfipair{i,va,vb,vc,vd,o}
		\fmfiequ{i}{(0,.5h)}
		\fmfiv{label=$\gamma_{8_{\pm}}$,label.angle=80,label.dist=1cm}{i}
		\fmfiequ{va}{(.3w,.5h)}
		\fmfiequ{vb}{(.5w,.8h)}
		\fmfiequ{vc}{(.7w,.5h)}
		\fmfiequ{vd}{(.5w,.2h)}
		\fmfiequ{o}{(w,.5h)}
		\fmfi{dashes}{i--va}
		\fmfi{dashes}{vc{right} .. {right}o}
		\fmfi{plain}{va{up} .. tension 1 .. {right}vb}
		\fmfi{plain}{vb{right} .. tension 1 .. {down}vc}
		\fmfi{plain}{vc{down} .. tension 1 .. {left}vd}
		\fmfi{plain}{vd{left} .. tension 1 .. {up}va}
		\fmfi{photon}{vb{down} .. {down}vd}
	\end{fmfgraph*}
	\quad
	\begin{fmfgraph*}(37,25)
		\fmfipair{i,va,vb,vc,vd,o}
		\fmfiequ{i}{(0,.5h)}
		\fmfiv{label=$\gamma_{9_{\pm}}$,label.angle=80,label.dist=1cm}{i}
		\fmfiequ{va}{(.3w,.5h)}
		\fmfiequ{vb}{(.5w,.8h)}
		\fmfiequ{vc}{(.7w,.5h)}
		\fmfiequ{vd}{(.5w,.2h)}
		\fmfiequ{o}{(w,.5h)}
		\fmfi{dashes}{i--va}
		\fmfi{dashes}{vc{right} .. {right}o}
		\fmfi{plain}{va{up} .. tension 1 .. {right}vb}
		\fmfi{plain}{vb{right} .. tension 1 .. {down}vc}
		\fmfi{plain}{vc{down} .. tension 1 .. {left}vd}
		\fmfi{plain}{vd{left} .. tension 1 .. {up}va}
		\fmfi{gluon}{vd{up} .. {up}vb}
	\end{fmfgraph*}
	\quad
	\begin{fmfgraph*}(37,25)
		\fmfipair{i,va,vb,vc,vd,o,c}		
		\fmfiequ{i}{(0,.5h)}
		\fmfiv{label=$\gamma_{10_{\pm}}$,label.angle=80,label.dist=1cm}{i}
		\fmfiequ{va}{(.3w,.5h)}
		\fmfiequ{vb}{(.5w,.8h)}
		\fmfiequ{vc}{(.7w,.5h)}
		\fmfiequ{vd}{(.5w,.2h)}
		\fmfiequ{o}{(w,.5h)}
		\fmfiequ{c}{(.5w,.5h)}
		\fmfi{dashes}{i--va}
		\fmfi{dashes}{vc{right} .. {right}o}
		\fmfi{plain}{va{up} .. tension 1 .. {right}vb}
		\fmfi{plain}{vb{right} .. tension 1 .. {down}vc}
		\fmfi{plain}{vc{down} .. tension 1 .. {left}vd}
		\fmfi{plain}{vd{left} .. tension 1 .. {up}va}
		\fmfi{photon}{vb{down} .. {down}c}
		\fmfi{gluon}{vd{up} .. {up}c}
	\end{fmfgraph*}
	\end{fmffile} 
\end{figure}
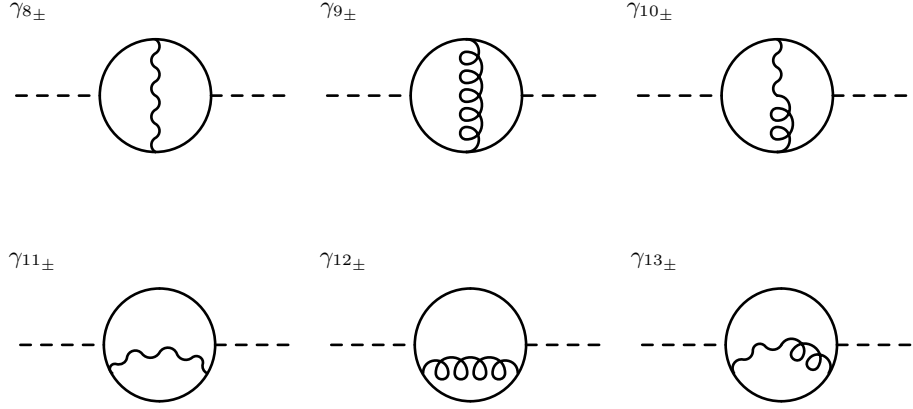
\begin{figure}[h]
\center
  \begin{fmffile}{2loops}
  	\begin{fmfgraph*}(37,25)
	  	\fmfipair{i,va,vb,vc,vd,o,c,ve,vf}		
		\fmfiequ{i}{(0,.5h)}
		\fmfiv{label=$\gamma_{11_{\pm}}$,label.angle=80,label.dist=1cm}{i}
		\fmfiequ{va}{(.3w,.5h)}
		\fmfiequ{vb}{(.5w,.8h)}
		\fmfiequ{vc}{(.7w,.5h)}
		\fmfiequ{vd}{(.5w,.2h)}
		\fmfiequ{o}{(w,.5h)}
		\fmfiequ{c}{(.5w,.5h)}
		\fmfiequ{ve}{c+(-.17w,-.15h)}
		\fmfiequ{vf}{c+(.17w,-.15h)}
		\fmfi{photon}{ve{up} .. tension 2.3 .. {down}vf}
		\fmfi{dashes}{i--va}
		\fmfi{dashes}{vc{right} .. {right}o}
		\fmfi{plain}{va{up} .. tension 1 .. {right}vb}
		\fmfi{plain}{vb{right} .. tension 1 .. {down}vc}
		\fmfi{plain}{vc{down} .. tension 1 .. {left}vd}
		\fmfi{plain}{vd{left} .. tension 1 .. {up}va}
	\end{fmfgraph*}	
	\quad
	  	\begin{fmfgraph*}(37,25)
	  	\fmfipair{i,va,vb,vc,vd,o,c,ve,vf}		
		\fmfiequ{i}{(0,.5h)}
		\fmfiv{label=$\gamma_{12_{\pm}}$,label.angle=80,label.dist=1cm}{i}
		\fmfiequ{va}{(.3w,.5h)}
		\fmfiequ{vb}{(.5w,.8h)}
		\fmfiequ{vc}{(.7w,.5h)}
		\fmfiequ{vd}{(.5w,.2h)}
		\fmfiequ{o}{(w,.5h)}
		\fmfiequ{c}{(.5w,.5h)}
		\fmfiequ{ve}{c+(-.17w,-.15h)}
		\fmfiequ{vf}{c+(.17w,-.15h)}
		\fmfi{curly}{vf .. tension 1 .. ve}
		\fmfi{dashes}{i--va}
		\fmfi{dashes}{vc{right} .. {right}o}
		\fmfi{plain}{va{up} .. tension 1 .. {right}vb}
		\fmfi{plain}{vb{right} .. tension 1 .. {down}vc}
		\fmfi{plain}{vc{down} .. tension 1 .. {left}vd}
		\fmfi{plain}{vd{left} .. tension 1 .. {up}va}
	\end{fmfgraph*}	
	\quad
	 \begin{fmfgraph*}(37,25)
	  	\fmfipair{i,va,vb,vc,vd,o,c,ve,vf}		
		\fmfiequ{i}{(0,.5h)}
		\fmfiv{label=$\gamma_{13_{\pm}}$,label.angle=80,label.dist=1cm}{i}
		\fmfiequ{va}{(.3w,.5h)}
		\fmfiequ{vb}{(.5w,.8h)}
		\fmfiequ{vc}{(.7w,.5h)}
		\fmfiequ{vd}{(.5w,.2h)}
		\fmfiequ{o}{(w,.5h)}
		\fmfiequ{c}{(.5w,.5h)}
		\fmfiequ{ve}{c+(-.17w,-.15h)}
		\fmfiequ{vf}{c+(.17w,-.15h)}
		\fmfi{curly}{vf--c}
		\fmfi{photon}{c{left} .. tension .8 .. ve}
		\fmfi{dashes}{i--va}
		\fmfi{dashes}{vc{right} .. {right}o}
		\fmfi{plain}{va{up} .. tension 1 .. {right}vb}
		\fmfi{plain}{vb{right} .. tension 1 .. {down}vc}
		\fmfi{plain}{vc{down} .. tension 1 .. {left}vd}
		\fmfi{plain}{vd{left} .. tension 1 .. {up}va}
	\end{fmfgraph*}	
	\caption{The 2-loops vacuum-polarization tensor graphs, which the continuous lines represent propagators of either $\j_+$ or $\j_-$, and the dashed lines the external legs of either $A_\m$ or $a_\m$.}\label{2loopgraphs}
  \end{fmffile}
\end{figure}

Complementary to the previous dimensional discussion, a tensor structure analysis of the 2-loops vacuum-polarization tensor integrands is opportune. First of all, the thirty six 2-loops vacuum-polarization tensors diagrams ($\gamma_{i_{\pm}}$, $i=8\dots13$) are displayed in Fig. \ref{2loopgraphs}, and their UV superficial degrees of divergence ($d(\gamma_{i_{\pm}})$) are $d(\gamma_{8_{\pm}})=d(\gamma_{9_{\pm}})=d(\gamma_{11_{\pm}})=d(\gamma_{12_{\pm}})=0$ and $d(\gamma_{10_{\pm}})=d(\gamma_{13_{\pm}})=-1$, thus from the former UV degree of divergences, the graphs $\gamma_{8_{\pm}}$, $\gamma_{9_{\pm}}$, $\gamma_{11_{\pm}}$ and $\gamma_{12_{\pm}}$ have to be renormalized, on the other hand the graphs $\gamma_{10_{\pm}}$ and $\gamma_{13_{\pm}}$ are already UV finite. Also, prior to the proof on the non generation of possible parity-odd Levi-Civita symbol dependent counterterms, it is suitable to write down explicitly the divergent vacuum-polarization tensors corresponding to the diagrams\footnote{As already mentioned, since possible parity-even local counterterm of the type $\epsilon^{\mu \a \nu}A_{\mu}p_{\a}a_{\nu}$ has not been taken into consideration, it remains sixteen graphs that could generate parity-odd-like counterterms $\epsilon^{\mu \a \nu}A_{\mu}p_{\a}A_{\nu}$ and $\epsilon^{\mu \a \nu}a_{\mu}p_{\a}a_{\nu}$.} $\gamma_{8_{\pm}}$, $\gamma_{9_{\pm}}$, $\gamma_{11_{\pm}}$ and $\gamma_{12_{\pm}}$:
\ba  
&&\P_{\gamma_{8_{\pm}}}^{\m\n}(p,s)=\l_8^2\int{\frac{d^3k_1}{(2\p)^3}}\int{\frac{d^3k_2}{(2\p)^3}}~e^2~{\wh I}_{\pm}^{\m\n}(k_1,k_2,p,s)~, \\
&&\P_{\gamma_{9_{\pm}}}^{\m\n}(p,s)=\l_9^2\int{\frac{d^3k_1}{(2\p)^3}}\int{\frac{d^3k_2}{(2\p)^3}}~g^2~{\wh I}_{\pm}^{\m\n}(k_1,k_2,p,s)~;
\ea
and 
\ba  
&&\P_{\gamma_{11_{\pm}}}^{\m\n}(p,s)=\l_{11}^2\int{\frac{d^3k_1}{(2\p)^3}}\int{\frac{d^3k_2}{(2\p)^3}}~e^2~{\wt I}_{\pm}^{\m\n}(k_1,k_2,p,s)~, \\
&&\P_{\gamma_{12_{\pm}}}^{\m\n}(p,s)=\l_{12}^2\int{\frac{d^3k_1}{(2\p)^3}}\int{\frac{d^3k_2}{(2\p)^3}}~g^2~{\wt I}_{\pm}^{\m\n}(k_1,k_2,p,s)~;
\ea
such that $\l_i=e$ ($i=8,9,11,12$) if the two external legs are of $A_\m$, otherwise, if $a_\m$ as the two external legs, $\l_i=g$, and 
\begin{multline}
	{\wh I}_{\pm}^{\m\n}(k_1,k_2,p,s)=-{\rm Tr}\left\{
	\gamma^{\mu} 
	\underbrace{
	\left[ i \frac{\sl k_1\mp m(s-1)}{k_1^2-m^2(s-1)^2} 
	\right]}
	\gamma_{\alpha}
	\left[-i\frac{1}{(k_1-k_2)^2-\mu^2}
	\left(\eta^{\alpha \beta}-\frac{(k_1^{\alpha}-k_2^{\alpha})(k_1^{\beta}-
	k_2^{\beta})}{(k_1-k_2)^2}\right)\right]\times\right. \\
	\times\left.\underbrace{\left[ i \frac{\sl k_2\mp m(s-1)}{k_2^2-m^2(s-1)^2} \right]}
	\gamma^{\nu}
	\underbrace{
	\left[ i \frac{(\sl k_2- \sl p)\mp m(s-1)}{(k_2-p)^2-m^2(s-1)^2} \right]}
	\gamma_{\beta}
	\underbrace{
	\left[ i \frac{(\sl k_1- \sl p)\mp m(s-1)}{(k_1-p)^2-m^2(s-1)^2} \right]}	
	\right\}~, \label{I-tilde}
\end{multline}
\begin{multline}
  	{\wt I}_{\pm}^{\m\n}(k_1,k_2,p,s)=-{\rm Tr}\left\{
	\gamma^{\mu} 
	\underbrace{
	\left[ i \frac{\sl k_1\mp m(s-1)}{k_1^2-m^2(s-1)^2} 
	\right]}
	\gamma^{\n}
	\underbrace{
	\left[ i \frac{(\sl k_1- \sl p)\mp m(s-1)}{(k_1-p)^2-m^2(s-1)^2} \right]}
	\g_\a
	\right.
	\left[-i\frac{1}{k_2^2-\mu^2}
	\left(\eta^{\alpha \beta}-\frac{k_2^{\alpha}k_2^{\beta}}
	{k_2^2}\right)\right]\times \\
	\times\left.\underbrace{\left[ i \frac{(\sl k_1-\sl k_2-\sl p)\mp m(s-1)}{(k_1-k_2-p)^2-m^2(s-1)^2} \right]}
	\gamma_{\b}
	\underbrace{
	\left[ i \frac{(\sl k_1- \sl p)\mp m(s-1)}{(k_1-p)^2-m^2(s-1)^2} \right]}
	\right\}~,\label{I-hat}
\end{multline}
where $p$ is the external momentum and the subscripts $+$ and $-$ refer to the internal lines of $\j_+$ and $\j_-$, respectively. 

Drawing attention to the integrands ${\wt I}_{\pm}^{\m\n}$ (\ref{I-tilde}) and ${\wt I}_{\pm}^{\m\n}$ (\ref{I-hat}), it can be seen that trace of the product of four to eight gamma matrices is generated, notwithstanding that solely trace of five and seven gamma matrices produces the Levi-Civita symbol $\epsilon^{\mu\nu\rho}$ (\ref{identities}). Also, it shall be noticed from the terms of the integrands, ${\wt I}_{\pm}^{\m\n}$ (\ref{I-tilde}) and ${\wt I}_{\pm}^{\m\n}$ (\ref{I-hat}), identified by under braces that they contribute each one to the trace product with at most one gamma matrix. Furthermore, as an example, by picking out from the integrand ${\wt I}_{\pm}^{\m\n}$ (\ref{I-tilde}) a piece of trace product of five gamma matrices, {\it e.g.}: 
$\cz_{5\pm}^{\m\n}(k_1,k_2,p,s) = -{\rm Tr} \{\g^\m [\mp im(s-1)] \g_\a [\D^{\a\b}(k_1,k_2)] [\mp im(s-1)] \g^\n [\mp im(s-1)] \g_\b [i(\sl k_1- \sl p)]\}$, it can be written as $\cz_{5\pm}^{\m\n}(k_1,k_2,p,s) = \pm \e^{\m\n\r} \cx_{5\r}(k_1,k_2,p,s) + \cy_{5\pm}^{\m\n}(k_1,k_2,p,s)$, where the first term is parity-odd whereas the second one is parity-even. In the sequence, using the same strategy applied to all five gamma matrices dependent terms, of the integrands (\ref{I-tilde}) and (\ref{I-hat}), they can be rewritten as:
\ba
&&{\wh I}_{5\pm}^{\m\n}(k_1,k_2,p,s) = \pm \e^{\m\n\r} {\wh \ca}_{5\r}(k_1,k_2,p,s) + {\wh \cs}_{5\pm}^{\m\n}(k_1,k_2,p,s)~,
\\
&&{\wt I}_{5\pm}^{\m\n}(k_1,k_2,p,s) = \pm \e^{\m\n\r} {\wt \ca}_{5\r}(k_1,k_2,p,s) + {\wt \cs}_{5\pm}^{\m\n}(k_1,k_2,p,s)~,
\ea
where ${\wh \cs}_{5\pm}^{\m\n}$ and ${\wt \cs}_{5\pm}^{\m\n}$ are parity-even tensors, and their subscripts $+$ and $-$ refer to the internal lines of $\j_+$ and $\j_-$, respectively. Consequently, the total integrands stemming from the trace of five gamma matrices, ${\wh I}_5^{\m\n}={\wh I}_{5+}^{\m\n}+{\wh I}_{5-}^{\m\n}$ and ${\wt I}_5^{\m\n}={\wt I}_{5+}^{\m\n}+{\wt I}_{5-}^{\m\n}$, read:
\ba
&&{\wh I}_5^{\m\n}(k_1,k_2,p,s) = {\wh \cs}_{5+}^{\m\n}(k_1,k_2,p,s) + {\wh \cs}_{5-}^{\m\n}(k_1,k_2,p,s)~,
\\
&&{\wt I}_5^{\m\n}(k_1,k_2,p,s) = {\wt \cs}_{5+}^{\m\n}(k_1,k_2,p,s) + {\wt \cs}_{5-}^{\m\n}(k_1,k_2,p,s)~,
\ea
thence there is no Levi-Civita symbol $\epsilon^{\mu\nu\rho}$ dependent terms emerged from the trace of five gamma matrices contributing to the total divergent integrand of vacuum-polarization tensor, remaining therefore only parity-even terms. Beyond that, it lacks to discuss the issue of non generation of possible parity-odd Levi-Civita symbol dependent counterterms for the case of the trace product of seven gamma matrices. Analogously to the preceding discussion, from the integrands ${\wt I}_{\pm}^{\m\n}$ (\ref{I-tilde}) and ${\wt I}_{\pm}^{\m\n}$ (\ref{I-hat}), considering the terms highlighted by under braces, and for instance, by picking out from the integrand ${\wt I}_{\pm}^{\m\n}$ (\ref{I-tilde}) a piece of trace product of seven gamma matrices, {\it e.g.}: 
$\cz_{7\pm}^{\m\n}(k_1,k_2,p,s) = -{\rm Tr} \{\g^\m [i(\sl k_1)] \g_\a [\D^{\a\b}(k_1,k_2)] [i(\sl k_2)] \g^\n [i(\sl k_2- \sl p)] \g_\b [\mp im(s-1)]\}$, it follows that $\cz_{7\pm}^{\m\n}(k_1,k_2,p,s) = \pm \e^{\m\n\r} \cx_{7\r}(k_1,k_2,p,s) + \cy_{7\pm}^{\m\n}(k_1,k_2,p,s)$, with the first term being parity-odd whereas the second one being parity-even. In addition to, doing similarly to all seven gamma matrices dependent terms of (\ref{I-tilde}) and (\ref{I-hat}), it can be shown that:
\ba
&&{\wh I}_{7\pm}^{\m\n}(k_1,k_2,p,s) = \pm \e^{\m\n\r} {\wh \ca}_{7\r}(k_1,k_2,p,s) + {\wh \cs}_{7\pm}^{\m\n}(k_1,k_2,p,s)~,
\\
&&{\wt I}_{7\pm}^{\m\n}(k_1,k_2,p,s) = \pm \e^{\m\n\r} {\wt \ca}_{7\r}(k_1,k_2,p,s) + {\wt \cs}_{7\pm}^{\m\n}(k_1,k_2,p,s)~,
\ea
where ${\wh \cs}_{7\pm}^{\m\n}$ and ${\wt \cs}_{7\pm}^{\m\n}$ are parity-even tensors, and 
the internal lines of $\j_+$ and $\j_-$ in the corresponding graphs are respectively represented by the subscripts $+$ and $-$. Morover, from the trace of seven gamma matrices, the total integrands, ${\wh I}_7^{\m\n}={\wh I}_{7+}^{\m\n}+{\wh I}_{7-}^{\m\n}$ and ${\wt I}_7^{\m\n}={\wt I}_{7+}^{\m\n}+{\wt I}_{7-}^{\m\n}$, are given by:
\ba
&&{\wh I}_7^{\m\n}(k_1,k_2,p,s) = {\wh \cs}_{7+}^{\m\n}(k_1,k_2,p,s) + {\wh \cs}_{7-}^{\m\n}(k_1,k_2,p,s)~,
\\
&&{\wt I}_7^{\m\n}(k_1,k_2,p,s) = {\wt \cs}_{7+}^{\m\n}(k_1,k_2,p,s) + {\wt \cs}_{7-}^{\m\n}(k_1,k_2,p,s)~,
\ea
thus likewise the five gamma matrices case, there is no Levi-Civita symbol $\epsilon^{\mu\nu\rho}$ dependent terms yielded from the trace of seven gamma matrices, surviving only parity-even terms which contribute to the total divergent vacuum-polarization tensor.

Ultimately, based on the argumentations above, the 2-loops unsubtracted integrands associated to the vacuum-polarization tensors $\P_{AA}^{\m\n}$ and $\P_{aa}^{\m\n}$ do not produce parity-violating counterterms of the type, $\epsilon^{\mu \a \nu}A_{\mu}\pa_{\a}A_{\nu}$ and $\epsilon^{\mu \a \nu}a_{\mu}\pa_{\a}a_{\nu}$, therefore it is concluded that parity is still preserved at 2-loops under the BPHZL renormalization procedures. Besides, due to the fact that the UV divergences are restricted up to 2-loops, thus for higher perturbative orders greater than two there is no need of UV subtractions, consequently it is definitely proved that the BPHZL renormalization method preserves parity for the massless parity-even $U(1)\times U(1)$ Maxwell-Chern-Simons QED$_3$ model \cite{masslessU1U1QED3}.

\section{Conclusion}
\label{V}


The massless parity-even $U(1)\times U(1)$ planar quantum electrodynamics (QED$_3$) model \cite{masslessU1U1QED3} exhibits quantum parity conservation at all orders in perturbation theory. The proof has been performed using the Bogoliubov-Parasiuk-Hepp-Zimmermann-Lowenstein (BPHZL) renormalization method, however owing to the presence of two massless fermions in the spectrum, infrared divergences might emerge in the course of the ultraviolet divergences subtractions and must be subtracted as well, for this reason, the Lowenstein-Zimmermann (LZ) subtraction scheme has been adopted. The power-counting -- the ultraviolet and infrared superficial degrees of divergence (\ref{power_counting}) of any 1-particle irreducible Feynman diagram -- reveals that ultraviolet divergences are bounded at most to two loops. At one loop all six vacuum-polarization tensor diagrams are linear ultraviolet divergent, four of the six self-energy diagrams are logarithm ultraviolet divergent, while all the vertex-function diagrams are ultraviolet finite, beyond that at two loops, twenty four of the thirty six vacuum-polarization tensor Feynman graphs are ultraviolet divergent (Fig. \ref{1loopgraphs} and Table \ref{table2}). Although there are counterterms\footnote{The explicitly BPHZL renormalization and the calculations of all counterterms at 1- and 2-loops, whether parity-even or -odd, are left to another work \cite{BPHZL_masslessU1U1QED3}, since the purpose of this one was to verify if the LZ subtraction scheme in the framework of the BPHZ renormalization method would preserve or not parity symmetry.} at one and two loops, none of them violate parity symmetry and together to the fact that the model is superrenormalizable, it stems as a byproduct that parity is guaranteed at any radiative order. As a final conclusion, for the model presented in this work, opposite to the case of the ordinary massless parity-even $U(1)$ QED$_3$ \cite{massless-1-loop}, the BPHZL subtraction scheme with the Lowenstein's adaptation of the Zimmermann's forest formula \cite{lz} preserves parity symmetry at all perturbative order.

\subsection*{Acknowledgements}

O.M.D.C. dedicates this work to his father (Oswaldo Del Cima, {\it in memoriam}), mother (Victoria M. Del Cima, {\it in memoriam}), daughter (Vittoria), son (Enzo) and Glaura Bensabat. CAPES-Brazil is acknowledged for invaluable financial help.


\begin{references}

\bibitem{deser-jackiw-templeton-schonfeld} J.F. Schonfeld, Nucl. Phys. B185 (1981) 157; R. Jackiw and S. Templeton, Phys. Rev. D23 (1981) 2291; S. Deser, R. Jackiw and S. Templeton, Ann. Phys. (NY) 140 (1982) 372, Phys. Rev. Lett. 48 (1982) 975 and Ann. Phys. (NY) 281 (2000) 409. 

\bibitem{high-Tc} M. Franz, Z. Te\v{s}anovi\'c and O. Vafek, Phys. Rev. B66 (2002) 054535; I.F. Herbut, Phys. Rev. B66 (2002) 094504; H.R. Christiansen, O.M. Del Cima, M.M. Ferreira Jr and J.A. Helay\"el-Neto, Int. J. Mod. Phys. A18 (2003) 725.

\bibitem{quantum-hall-effect} R.B. Laughlin, Phys. Rev. Lett. 50 (1983) 1395; A.M.J. Schakel, Phys. Rev. D43 (1991) 1428; A. Raya and E.D. Reyes, J. Phys. A: Math. Theor. 41 (2008) 355401.
 
\bibitem{graphene} V.P. Gusynin, V.A. Miransky and I.A. Shovkovy, Phys. Rev. Lett. 73 (1994) 3499; K.S. Novoselov, A.K. Geim, S.V. Morozov, D. Jiang, M.I. Katsnelson, I.V. Grigorieva, S.V. Dubonos, A.A. Firsov,  Nature 438 (2005) 197; V.P. Gusynin, S.G. Sharapov and J.P. Carbotte, Int. J. Mod. Phys. B21 (2007) 4611; R. Jackiw and S.-Y. Pi, Phys. Rev. Lett. 98 (2007) 266402; M.I. Katsnelson and K.S. Novoselov, Solid State Comm. 143 (2007) 3; A.H. Castro Neto, F. Guinea, N.M.R. Peres, K.S. Novoselov and A.K. Geim, Rev. Mod. Phys. 81 (2009) 109; E.M.C. Abreu, M.A. De Andrade, L.P.G. De Assis, J.A. Helay\"el-Neto, A.L.M.A. Nogueira and R.C. Paschoal, J. High Energy Phys. 1105 (2011) 001; I. Fialkovsky and D.V. Vassilevich, Eur. Phys. J. B85 (2012) 384; O.M. Del Cima and E.S. Miranda, Eur. Phys. J. B91 (2018) 212.  
 
\bibitem{topological-insulators} M.Z. Hasan and C.L. Kane, Rev. Mod. Phys. 82 (2010) 3045; 
C.L. Kane and J.E. Moore, Physics World 24 (2011) 32; 
J.M. Fonseca, W.A. Moura-Melo and A.R. Pereira, J. Appl. Phys. 111 (2012) 064913. 
 
\bibitem{topological-superconductors} M. Leijnse and K. Flensberg, Semicond. Sci. Technol. 27 (2012) 124003; S. Yonezawa, arXiv:1604.07930v3, AAPPS Bulletin 26 No.3 (2016) 3; M. Sato and Y. Ando, Rep. Prog. Phys. 80 (2017) 076501. 

\bibitem{massless-all-orders} O.M. Del Cima, D.H.T. Franco and O. Piguet, Phys. Rev. D89 (2014) 065001.

\bibitem{massless-1-loop} O.M. Del Cima, D.H.T. Franco, O. Piguet and M. Schweda, Phys. Lett. B680 (2009) 108.  

\bibitem{epstein-glaser} P.H. De Moura, O.M. Del Cima, D.H.T. Franco, L.S. Lima and E.S. Miranda, ``No parity anomaly in massless QED$_3$: an Epstein-Glaser approach'', in preparation.

\bibitem{masslessU1U1QED3} W.B. De Lima, O.M. Del Cima and E.S. Miranda, Eur. Phys. J. B93  (2020) 187.

\bibitem{BPHZ} W. Zimmermann, Comm. Math. Phys. 15 (1969) 208 and \textit{Lectures on Elementary Particles and Quantum Field Theory}, 1970 Brandeis lectures, eds. S. Deser, M. Grisaru and H. Pendleton, MIT Press (Cambridge-USA), 1971.

\bibitem{lz} J.H. Lowenstein, Comm. Math. Phys. 47 (1976) 53; J.H. Lowenstein and W. Zimmermann, Nucl. Phys. B86 (1975) 77 and \textit{Renormalization Theory}, eds. G. Velo and A.S. Wightman, D. Reidel (Dordrecht-Holland), 1976.

\bibitem{Binegar} B. Binegar, J. Math. Phys. 23 (1982) 1511.

\bibitem{lautrup-nakanishi} N. Nakanishi, Progr. Theor. Phys. 35 (1966) 1111; Progr. Theor. Phys. 37 (1967) 618; B. Lautrup, Mat. Fys. Medd. Dan. Vid. Selsk 35 (1967) No.11.

\bibitem{brs} C. Becchi, A. Rouet and R. Stora, Comm. Math. Phys. 42 (1975) 127 and Ann. Phys. (N.Y.) 98 (1976) 287; O. Piguet and A. Rouet, Phys. Rep. 76 (1981) 1;  O. Piguet and S.P. Sorella, \textit{Algebraic Renormalization}, Lecture Notes in Physics, m28, Springer-Verlag (Berlin-Heidelberg), 1995.

\bibitem{BPHZL_masslessU1U1QED3} O.M. Del Cima, D.H.T. Franco, L.S. Lima and E.S. Miranda, ``The BPHZL renormalization of parity-preserving massless planar quantum electrodynamics'', in preparation.

\end{references}
\end{document}